# Evidence of orbit-selective electronic kagome lattice with planar flat-band in correlated paramagnetic $YCr_6Ge_6$

T. Y. Yang[1,#], Q. Wan[1,#], Y. H. Wang[2], M. Song[2], J. Tang[2], Z. W. Wang[3], H. Z. Lv[3], N. C. Plumb[4], M. Radovic[4], G. W. Wang[5], G. Y. Wang[6], Z. Sun[7], R. Yu[3], M. Shi[4], Y. M. Xiong[2,*] and N. Xu[1,*]

[1] *Institute of Advanced Studies, Wuhan University, Wuhan 430072, China*

[2] *Anhui Province Key Laboratory of Condensed Matter Physics at Extreme Conditions, High Magnetic Field Laboratory, Chinese Academy of Sciences, Hefei 230031, China*

[3] *School of Physics and Technology, Wuhan University, Wuhan 430072, China*

[4] *Swiss Light Source, Paul Scherrer Institut, CH-5232 Villigen PSI, Switzerland*

[5] *Analytical and Testing Center, Chongqing University, Chongqing 401331,China*

[6] *Chongqing Institute of Green and Intelligent Technology, Chinese Academy of Sciences, Chongqing 400714, China*

[7] *National Synchrotron Radiation Laboratory, University of Science and Technology of China, Hefei, Anhui 230029, China*

# These authors contributed equally to this work.

* E-mail: nxu@whu.edu.cn, yxiong@hmfl.ac.cn

**Electronic properties of kagome lattice have drawn great attention recently. In associate with flat-band induced by destructive interference and Dirac cone-type dispersion, abundant exotic phenomena have been theoretically discussed. The material realization of electronic kagome lattice is a crucial step towards comprehending kagome physics and achieving novel quantum phases. Here, combining angle-resolved photoemission spectroscopy, transport measurements and first-principle calculations, we expose a planar flat-band in paramagnetic $YCr_6Ge_6$ as a typical signature of electronic kagome lattice. We unearth that the planar flat-band arises from the $d_{z^2}$ electrons with intra-kagome-plane hopping forbidden by destructive interference. On the other hand, the destructive interference and flatness of the $d_{x^2-y^2}$ and $d_{xy}$ bands are decomposed possibly due to additional in-plane hopping terms, but the Dirac cone-type dispersion is reserved near chemical potential. We explicitly unveil that orbital character plays an essential role to realize electronic kagome lattice in bulk materials with transition metal kagome layers. Paramagnetic $YCr_6Ge_6$ provides an opportunity to comprehend intrinsic properties of electronic kagome lattice as well as its interplays with spin orbit coupling and electronic correlation of Cr-*3d* electrons, and be free from complications induced by strong local moment of ions in kagome planes.**

Due to the unique triangular-based corner sharing geometry, the kagome lattice (Fig. 1a) provides an exciting route for theoretical modelling and experimental seeking novel quantum phases. A kagome lattice system with inter-site antiferromagnetic coupling is predicted to present magnetic frustration, which could support a quantum spin liquid ground state [1-3]. The same lattice geometry with nearest-neighbour (NN) interaction can also create a completely destructive interference of electron wave functions, resulting in a quenching of itinerant electron's kinetic energy [4-6]. The quenched electrons are presented as non-dispersive flat-band in the momentum space (Fig. 1b), which are strongly correlated and can achieve exotic emergent phenomena if they sit close enough to Fermi energy ($E_F$), such as ferromagnetism [4-8], superconductivity [9] and Wigner crystal [10-12]. Similar as the graphene honeycomb lattice, Dirac cone-type dispersion (Dirac dispersion in short) exists in kagome lattice (Fig. 1b). Quantum anomalous Hall effect can be further realized, with the conditions that time-reversal symmetry is broken in the system and $E_F$ sits in the spin-orbit coupling (SOC) induced gap of the Dirac dispersion simultaneously [13-15]. Furthermore, the flat-band can be topological non-trivial with SOC and fractional quantum Hall effect could emerge at high temperature in kagome lattice if the flat-band is narrow enough [16-21].

From the experimental side, however, it is still challenging to realize electronic kagome lattice in bulk materials. Recently, systems with transition metals Mn [22-25], Fe [26-30] and Co [31-34] kagome planes in different stacking forms have been intensively studied. Evidences of novel electronic properties have been reported in these kagome lattice systems, such as large anomalous Hall effect [22-23, 25-27, 31-32] and negative magneto-resistance [24, 32-33] which may related to magnetic Weyl fermion. However, the electronic structures in these systems are relatively complicate and lack the one-to-one correspondence to that in electronic kagome lattice (Fig. 1b). Furthermore, the Mn, Fe and Co ions in kagome lattice systems have strong local moment with different type of exchange interactions. The various long-range magnetic orderings induce further complications for understanding these systems with first-principal calculations. Therefore, material realization of typical electronic kagome lattice is a crucial step to comprehend the intrinsic properties of the quenched electrons and Dirac dispersion, and further realize exotic quantum phases predicted in kagome lattice.

In this paper, combining angle-resolved photoemission spectroscopy (ARPES), transport experiments and first-principal calculations, we report a planar flat-band as a signature of electronic kagome lattice in $YCr_6Ge_6$, in which two crystalline equivalent Cr kagome planes are separated by two kinds of spacing layers ($YGe_2$ and $Ge_4$) in the unit cell (Fig. 1c). The non-distorted stacking of kagome layers results in a simple hexagonal bulk Brillouin zone (BZ) shown in Fig. 1d. In the $k_z = 0$ plane, the full set of finger prints of electronic kagome lattice is clearly observed for the $d_{z^2}$ bands, including flat-band, Dirac dispersion and saddle points. We reveal that interlayer interaction of $d_{z^2}$ electrons is non-negligible and makes the planar flat-band dispersive along the our-of-plane direction. The observation of planar flat-band manifests that the in-plane kinetic energy of $d_{z^2}$ electrons is quenched due to the destructive interference of kagome geometry, with that the intra-kagome-plane hopping is forbidden and the electrons can only hop between different kagome layers. The flat-band scenario is further supported by transport measurements with an abnormal resistivity anisotropy and relatively big Sommerfeld coefficient. For $d_{x^2-y^2}$ and $d_{xy}$ orbits, although the Dirac dispersion is suggested by our first-principal calculations and the corresponding lower branches are experimentally observed, the destructive interference and flat-band orbits are decomposed possibly due to the influence of additional in-kagome-plane interactions. Compare the experimental and calculated results, we found that the electrons in $YCr_6Ge_6$ are correlated with a mass enhancement of 1.6. SOC opens gaps at the Dirac points at the K point with gap size in tens of meV as indicated by calculations. Our work unambiguously uncovers that orbital character is important to format electronic kagome lattice in bulk materials with stacked transition metal kagome layers. Paramagnetic $YCr_6Ge_6$ with planar flat-band and Dirac dispersions offers a chance to investigate intrinsic properties of electronic kagome lattice and the interactions with SOC and electronic correlation, without complication induced by strong local moment of ions in kagome planes.

Firstly, we plot the photoemission intensity of $YCr_6Ge_6$ in a large binding energy ($E_B$) range in Fig. 2a, in that Cr *3p*, Ge *3d*/*3p* and Y *4p*/*3d* core level peaks are clearly observed. Note the double-peak features of Ge and Y core levels correspond to the different total angular momentum and no signature of surface contribution observed, suggesting the bulk origin of the photoemission signals. To explore the electronic states of Cr kagome plane in $YCr_6Ge_6$, we map the in-plane band structure with hν = 55 eV, corresponding to the $k_z$ = 0 plane, with the results shown in Fig. 2b-h. The Fermi surface (FS) shown in Fig. 2b consists of a big pocket (named as α) centered at the Γ point, and a hot spot (labeled as β) at the K point. The α pocket corresponds to a hole-like band crossing $E_F$, with Fermi momentum along the Γ-M and Γ-K directions indicated by arrows in Fig. 2c and d, respectively. Interestingly, additional flat-band, named as γ, is clearly observed near $E_F$, that shows no dispersion along the Γ-M and Γ-K directions (Fig. 2c and d). We note spectra weight of the α and flat γ bands in the first BZ are much less intense than that from higher BZs (Fig. 2b and c), due to the matrix element effect of photoemission. To eliminate the strong intensity variation and focus on the band dispersion, we plot the experimental results along the Γ-M direction in the $2^{nd}$ BZ (cut3 in Fig. 2b), with the ARPES intensity and energy distribution curve (EDC) plotted in Fig. 2e and f, respectively. The flat-band γ clearly appears with peaks near $E_F$ in EDC plot along the Γ-M direction between two $2^{nd}$ Γ points. Similarly, the flat-band γ is explicitly observed near $E_F$ along the Γ-K-M direction in higher BZs (cut4 in Fig. 2b), as seen from results shown in Fig. 2g and h.

The observation of nondispersive γ band along the high symmetry lines Γ-K-M-Γ is fully consistent with the electronic kagome lattice phase with electrons quenched by destructive interference. On the other hand, we note that the nondispersive band is also observed in other systems in a trivial form, for example in $BaCo_2As_2$ [35]. In the latter case, the flat-band is accidental and only appears along some specific momentum paths (such as the Γ-M direction in $BaCo_2As_2$). In $YCr_6Ge_6$, the flat-band not only appears along high symmetry lines, such as the Γ-K-M-Γ directions, but also shows no dispersion in whole the $k_z$ = 0 plane (see Supplementary Information Fig. S3). Our results exclude the trivial case of the nondispersive band and indicate an electronic kagome lattice origin of the flat-band γ.

In addition to the flat-band, the bulk electronic structure of $YCr_6Ge_6$ in the $k_z$ = 0 plane shows complete hallmarks of electronic kagome lattice including Dirac dispersions and saddle points. Figure 3a shows the energy evolution of electronic states in the $k_z = 0$ plane. As the energy goes from $E_F$ (Fig. 2b) to higher $E_B$ (Fig. 3a-I), the β band at K point evolves from a single hot spot to a circular pocket. In the meantime, the hole-like α pocket at the Γ points also becomes bigger. Further to higher $E_B$ (Fig. 3a-II), both the α and β pockets further expand and touch each other at $T_{\alpha\text{-}\beta}$ point along the Γ-K direction at $E_B \sim 0.26$ eV, that can also be clearly observed in ARPES intensity shown in Fig. 3b and curvature plots along the Γ-K-M-K-Γ direction shown in Fig. 3c. Then the α and β pockets become even bigger at higher $E_B$, and the two α bands touch each other at the $S_\alpha$ point along the Γ-M direction at $E_B \sim$ 0.4 eV (Fig. 3a-III), as also seen from the ARPES intensity and curvature plots along Γ-M-Γ direction shown in Fig. 3d and e, respectively. From the $S_\alpha$ point, the topology of the α band evolves from big hole-like pockets around different Γ points to small electron-like pockets around the K point. Note that the touching point at the $S_\alpha$ point is a saddle point, at which the effective mass m* along the Γ-M direction is positive and that along the M-K direction is negative. Then two β bands touch each other at the $S_\beta$ point along the M-K direction at $E_B = 0.5$ eV (Fig. 3b-e), that can also be clearly resolved from the constant energy plot in Fig. 3a-IV. The β band evolves from six small hole-like pockets at different K points to single big electron-like pockets centered at the Γ point as $E_B > 0.5$ eV, that becomes smaller and moves towards to the center of BZ with higher $E_B$, as seen from Fig. 3aIV-VI. Similar to $S_\alpha$, the touching points $S_\beta$ are also saddle points, with positive m* along the K-M direction and negative m* along the Γ-M direction. Furthermore, the electron-like α pocket at the K point shrinks into a single point $D_\alpha$ at $E_B \sim 0.6$ eV (Fig. 3a-V), and become a hole-like pockets at higher $E_B$ (Fig. 3a-VI). The Dirac cone dispersion with the Dirac point $D_\alpha$ at the $E_B \sim 0.6$ eV can also been resolved in band dispersion along the Γ-K-M direction in Fig. 3b-c. Here the Dirac point $D_\alpha$ is not protected by symmetry and SOC will open a gap. Our calculations discussed later (Fig. 4f) suggest the SOC induced gap is about 10 meV which cannot be resolved in our experimental data.

Both the experimentally determined FS evolution (Fig. 3a) and the extracted dispersions (Fig. 4a) of the α and γ bands agree well with a prototypical electronic kagome lattice phase shown in Fig. 1b. However, the predicted novel quantum phases of 2D electronic kagome lattice, including ferromagnetism and superconductivity, are not observed in $YCr_6Ge_6$ down to 2K [36, and Supplementary Information Fig. S2]. Our photon energy dependent ARPES measurements indicate that inter-kagome-layers interaction is non-negligible in $YCr_6Ge_6$ and makes the system deviate from 2D electronic kagome lattice. In Fig. 4b, we present the ARPES spectra at $E_F$ in the $k_x$-$k_z$ plane, measured with hν in a range of 20-120 eV. The periodical variation along the $k_z$ direction is a direct evidence of the bulk electronic states origins for the ARPES signals. We note that the periodicity along $k_z$ is of 4π/c, as twice as that in the bulk BZ along the $k_z$ direction (2π/c), because the two Cr kagome lattice in the unite cell are crystalline equivalence (Fig. 1c). Along the out of plane direction Γ-A-Γ, as seen from the intensity plot in Fig. 4c, ARPES intensity near $E_F$ only appears near the Γ point, corresponding to the flat-band γ closed to $E_F$ in the $k_z = 0$ plane. The photon energy dependent ARPES results demonstrate a planar flat-band nature of γ, which is nondispersive feature in the $k_z = 0$ plane, indeed dispersive along the $k_z$ direction.

Our first-principal calculations projected on Cr-$3d_{z^2}$ orbit (Fig. 4d) show a good agreement with the observed α and γ bands (Fig. 4a) and the electronic kagome lattice (Fig. 1b). Indeed, the pronounced $k_z$ dependence (Fig. 4c) implies the $d_{z^2}$ orbital character of the α and γ bands. Our calculations further suggest that the observed β band corresponds the lower branches of the Dirac dispersion for the Cr-$3d_{x^2-y^2}$ and $3d_{xy}$ states (Fig. 4g), consistent with the quasi-two dimensional behavior of the β band (Supplementary Information Fig. S5). As seen from the insets of Fig. 4d and g, SOC opens gaps around 10 and 30 meV for the Dirac points at the K point for the $d_{z^2}$ and $d_{x^2-y^2}/3d_{xy}$ states, respectively. There are other hole-like bands with $d_{x^2-y^2}/d_{xy}$ (Fig. 4h) and $d_{xz}/d_{yz}$ (Supplementary Information Fig. S6) orbital characters cross $E_F$, which are quite close to the α band. Some hints of the additional hole-like bands are resolved in ARPES results shown in Fig. 2b and e. We note that a renormalization fact of 1.6 is needed for the calculation results to fit the overall features of experimental results (Supplementary Information Fig. S4), indicating a moderate electronic correlation effect in $YCr_6Ge_6$.

The experimentally determined band structure and first principal calculations consistently indicate that the in-plane kinetic energy of $d_{z^2}$ electrons is quenched by destructive interference, resulting in the planar flat-band $\gamma$. The unique behavior of the planar flat-band is related to the intrinsic orbital character for the layered kagome lattice. As seen from Fig. 4e-f, the $d_{z^2}$ states can only form $\pi$-bond in the kagome plane. The intra-kagome-plane hopping process is dominated by the NN interaction ($t_1$) and other in-plane hopping terms ($t_2$ and $t_3$) are negligibly weak due to the nature of $\pi$-bond. Therefore, both the experimentally determined and calculated $d_{z^2}$ bands in the $k_z$ = 0 plane (Fig. 4a and d) show a good agreement with the 2D electronic kagome lattice (Fig. 1b), with nondispersive band $\gamma$ induced by destructive interference of wavefunctions within the Cr kagome plane (Fig. 1a) and Dirac dispersion. On the other hand, the $d_{z^2}$ orbitals between the adjacent kagome planes can form effective $\sigma$ bonds ($t_4$ and $t_5$ in Fig. 4f), intermediated by the spacing layers (Fig. 1c). The inter-kagome-plane hoping terms lead to abnormal planar flat-band behavior of the $\gamma$ band near $E_F$, whose effective mass ($m^*$) is infinitely large in the $k_z$ = 0 plane but with finite value along the $k_z$ direction. The observation of planar flat-band manifests that the kinetic energy of $d_{z^2}$ electrons are planar quenched, of which the hopping within the Cr-kagome plane is forbidden and electrons can only move between different kagome planes. The abnormal transport anisotropy in $YCr_6Ge_6$ (Fig. 4j), with the in-plane resistivity more than twice larger than the out-of-plane one, provides a transport signature of the planar flat-band.

The planar flat-band leads to a peak in DOS near $E_F$, which is also suggested by our calculations shown in Fig. 4k. Correspondingly, a relatively big Sommerfeld coefficient is expected due to the contribution of the planar flat-band. The experimental Sommerfeld coefficient in our samples is estimated to be 80.5 mJ $K^{-2}$ $mol^{-1}$, by fitting our heat capacity data down to 2 K (Fig. 4l). The relatively big Sommerfeld coefficient experimentally determined by heat capacity measurements serves as another transport property supporting the planar flat-band, in addition to the unusual resistivity anisotropy.

However, comparing to the DOS anomaly in 2D kagome lattice, the hopping channels between different kagome layers for the planar flat-band relax the DOS peak near $E_F$. The suppression of the DOS near $E_F$ could be the reason for the absences of novel quantum phenomena in $YCr_6Ge_6$ down to 2 K, e.g. flat-band ferromagnetism and superconductivity.

Distinct from the $d_{z^2}$ electrons, the $d_{x^2-y^2}$ and $d_{xy}$ orbits in different kagome plane can only form δ bond (Fig. 4i). The corresponding inter-layer interaction is expected to be weak, which results in less dispersive feature along the Γ-A direction for $d_{x^2-y^2}$ and $d_{xy}$ states (Fig. 4g). The weak $k_z$ dependence of the $d_{x^2-y^2}$ and $d_{xy}$ states is experimentally confirmed by the quasi-two dimensional feature of the β band (Supplementary Information Fig. S5). However, the electronic kagome lattice behavior, namely the flat-band induced by destructive interference, is not observed in the first-principal calculation of the $d_{x^2-y^2}$ and $d_{xy}$ orbits (Fig. 4g). We note there are peak-like features in the DOS plot from 0.5 eV to 1 eV above $E_F$ for the $d_{x^2-y^2}$ and $d_{xy}$ orbits (Fig. 4k), which could be related to some heavy bands. However, the corresponding bands have a half eV in-plane band-width (Fig. 4g), indicating the fail of destructive interference in the kagome lattice. As seen from Fig. 4h, the $d_{x^2-y^2}$ and $d_{xy}$ orbits lay in the kagome plane and can form additional in-plane bonds between the second and third nearest-neighbour sites with strong σ and π components, respectively. It can lead to reasonable values of the additional in-plane hopping terms $t_2$ and $t_3$, which could destroy the destructive interference in the kagome lattice and give rise to a finite band-width.

In summary, we directly observed a planar flat-band and Dirac dispersion in paramagnetic $YCr_6Ge_6$ by ARPES, as signatures of electronic kagome lattice. The consistent ARPES and first-principal calculations indicate that the planar flat-band arises from the in-plane kinetic energy quenched $d_{z^2}$ electrons, with intra-kagome-plane hopping forbidden by destructive interference and hopping only allowed between different kagome planes. The abnormal resistivity anisotropy and relatively big Sommerfeld coefficient in transport measurements support the flat-band scenario. On the other hand, the destructive interference and flatness of the $d_{x^2-y^2}$ and $d_{xy}$ bands are decomposed possibly by additional in-plane interactions, but the Dirac cone-type dispersion is reserved near chemical potential with a small SOC induced gap. Our results demonstrate that orbital character plays an essential role for realization of electronic kagome lattice in bulk materials with transition metal kagome layers. Interestingly, the experimental and calculated results indicate the coexistence of reasonable SOC and electronic correlation effect of the Cr-*3d* electrons in $YCr_6Ge_6$.

Paramagnetic $YCr_6Ge_6$ with planar flat-band and Dirac dispersions provides a platform to investigate intrinsic properties of electronic kagome lattice and its interactions with SOC and electronic correlation, without complications induced by strong local moment of ions in kagome planes.

## Method

High-quality $YCr_6Ge_6$ single crystals were grown by the flux method using tin as a flux. The starting elements Y: Cr: Ge: Sn = 1: $\frac{7}{6}$: 6: 20 with purity higher than 99.9% were placed in an alumina crucible inside an evacuated quartz tube. The 5g mixture was heated to 1100 °C over 8 h, kept at 1100 °C for 10 h and then slowly cooled to 600 °C at the rate of 3 °C/h, and finally decanted in a centrifuge. The product was immersed in hydrochloric acid and smoothed by sanding in order to remove the tin flux, and then shiny single crystals were obtained. Clean surfaces for ARPES measurements were obtained by cleaving $YCr_6Ge_6$ samples in situ in a vacuum better than $5 \times 10^{-11}$ Torr. ARPES measurements were performed the "surface and interface" beamline (SIS) of the Swiss Light Source (SLS) with a Scienta Omicron R4000 analyzer, with an overall energy resolution of the order of 20 meV and angular resolution of 0.1°. The electronic structure of $YCr_6Ge_6$ was calculated based on the density functional theory (DFT) and the local density approximation (LDA) for the exchange correlation potential, as implemented in the plane-wave pseudopotential based Vienna ab initio simulation package (VASP). The wave functions were expanded in plane waves with a cutoff energy of 470 eV and Monkhorst–Pack *k* points were 9×9×5. The residual forces are less than 0.01eV/Å and SOC is included by using the second-order variational procedure.

## Figures

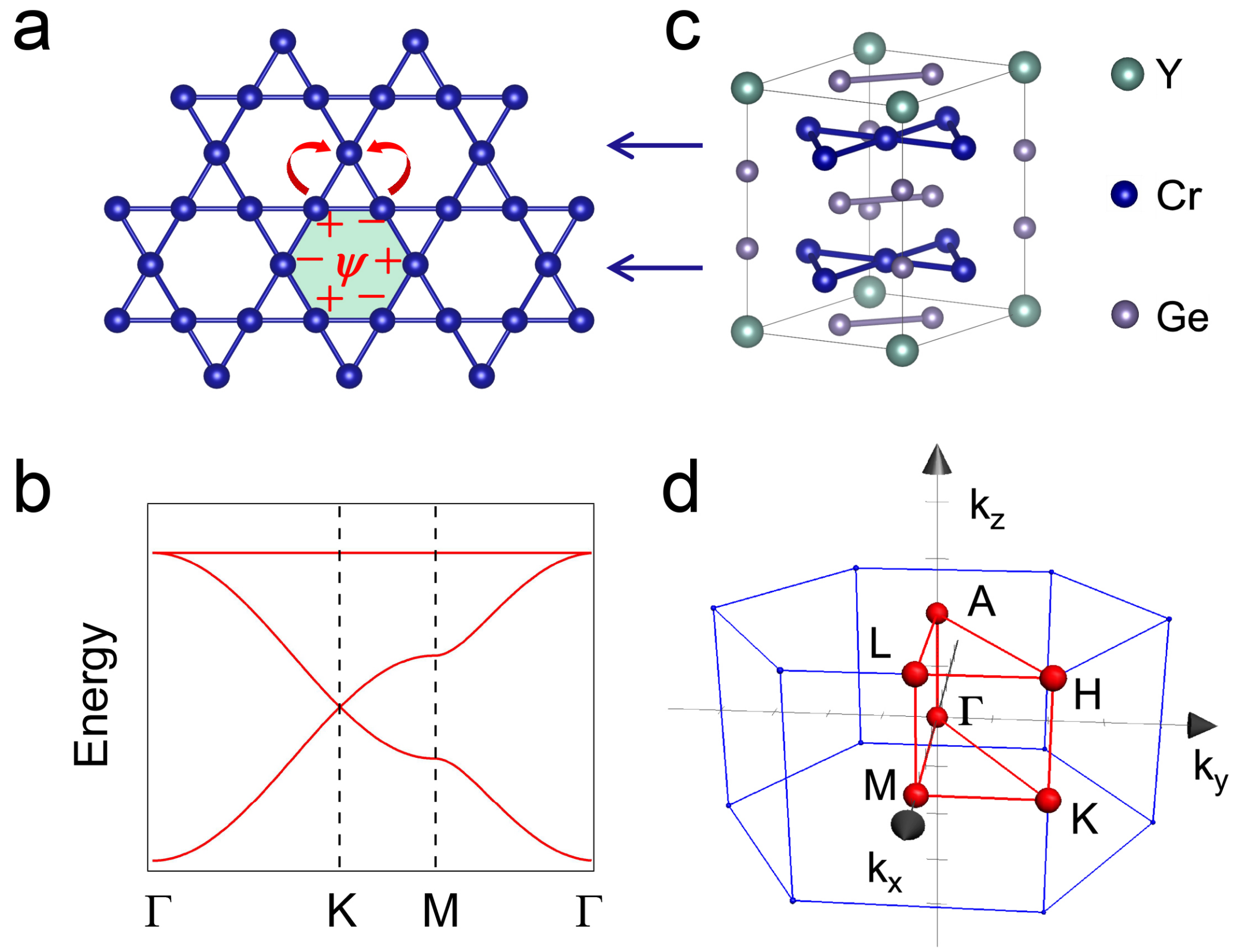

**Figure 1. Crystal and electronic structures of kagome lattice. a,** structure of kagome lattice and a quenched eigenstate induced by destructive interference. b, The band structure of electronic kagome lattice, with NN interaction dominating the in-plane hopping process. **c,** Crystal structure of $YCr_6Ge_6$, with Cr-kagome planes in a non-distorted stacking form. **d,** Bulk Brillouin zone of $YCr_6Ge_6$ with the high-symmetry points indicated.

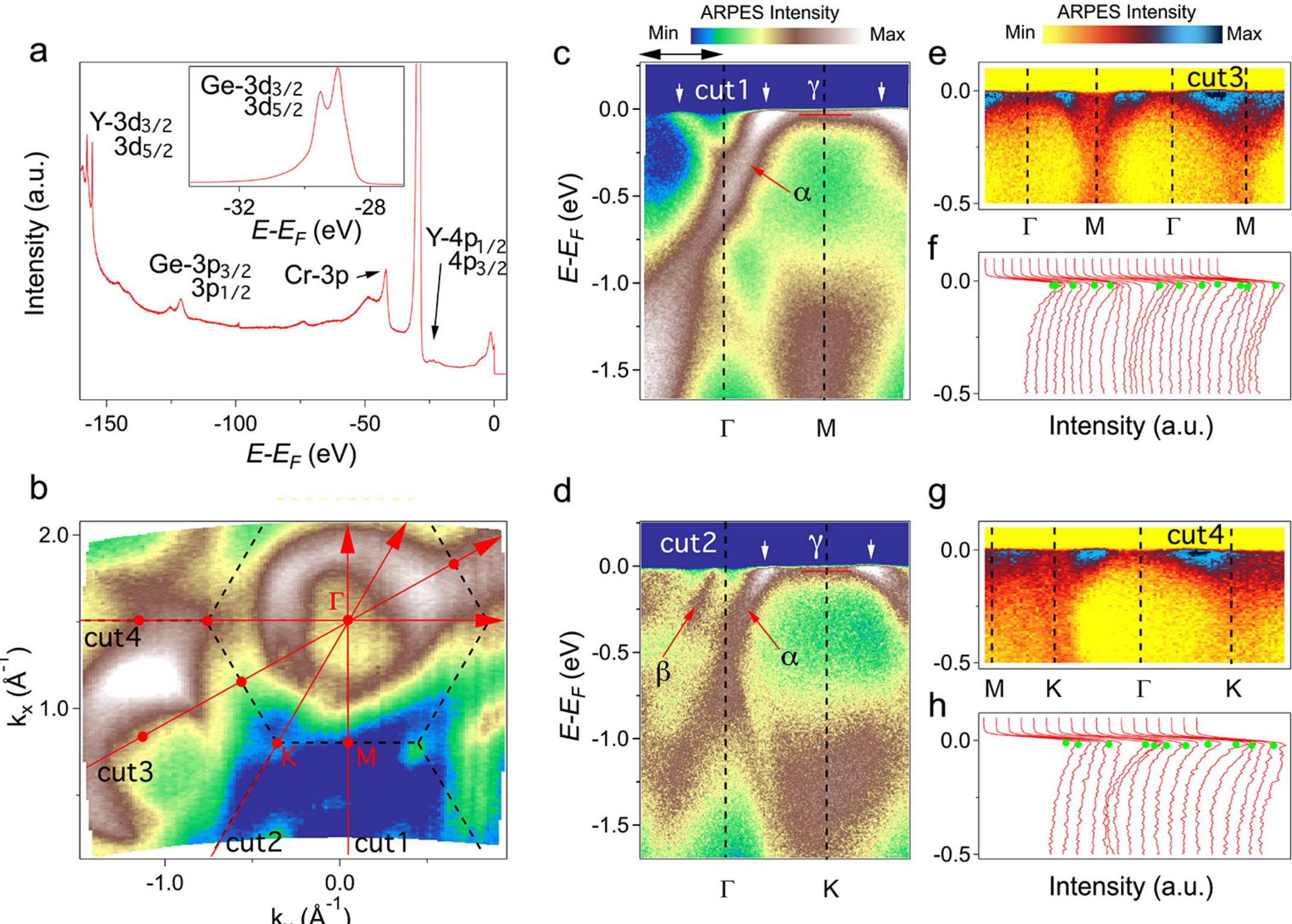

**Figure 2. Flat-band in the $k_z$ = 0 plane hosted in $YCr_6Ge_6$. a,** Core-level spectra taken with hν = 195 eV. The total angular momentum quantum numbers are labelled for peaks. **b,** Photoemission intensity plot at $E_F$ in the $k_z$ = 0 plane. **c-d,** Photoemission intensity plot along the Γ-M and Γ-K directions, with the momentum path indicated as cut1 and cut2 in **b**, respectively. **e-f**. Photoemission intensity and EDC plots along the cut3 direction in **b**, with the flat band peak marked. **g-h,** Same as **e-f**, but along the cut4 direction in **b**.

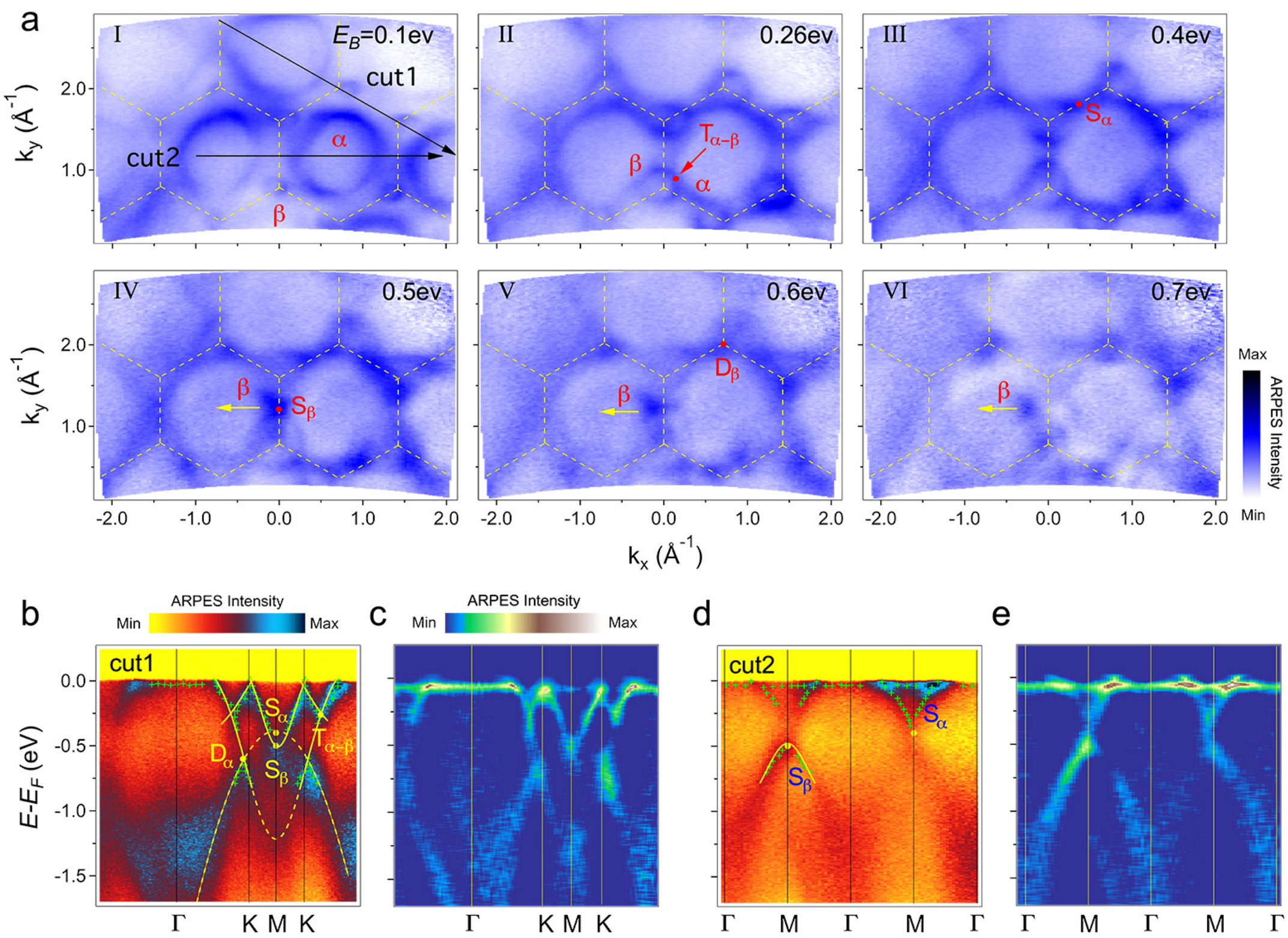

**Figure 3. Band structure evolution in the $k_z$ = 0 plane. a,** Photoemission intensity plots at different constant $E_B$ in the $k_x$-$k_y$ plane. **b,** Photoemission intensity plot along the Γ-K-M-K-Γ direction, with the extracted band dispersions (crosses) and guiding lines overlaid on top. **c,** The corresponding curvature plot of the spectrum in **b**. **d-e,** Same as **b-c,** but along the Γ-M-Γ direction. The touching points and saddle points are labelled in **a**, **b** and **d**.

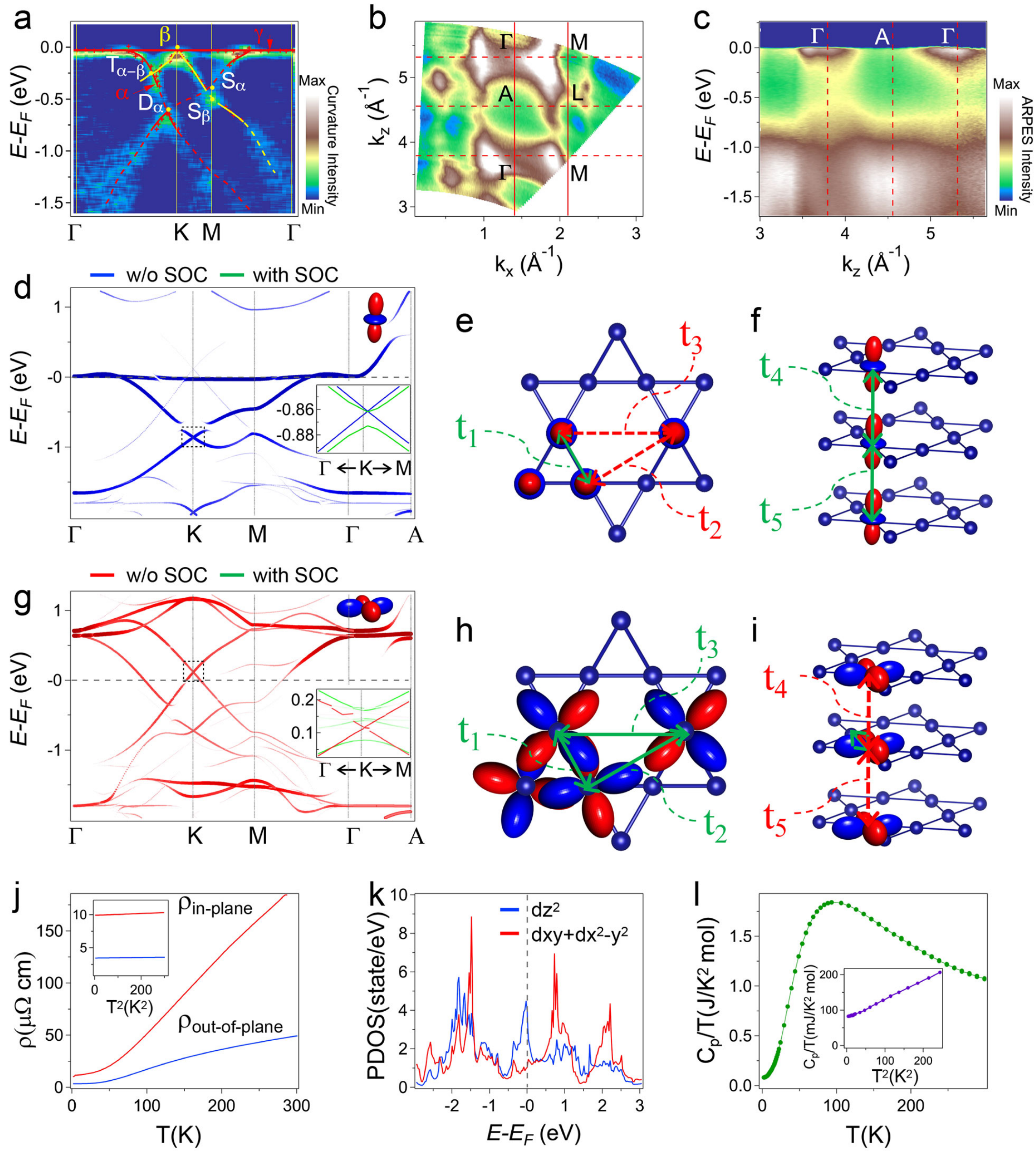

**Figure 4. The orbital-selective electronic kagome lattice in $YCr_6Ge_6$. a,** The curvature plot along the Γ-K-M-Γ direction. The extracted band dispersions and the guiding lines are overlaid on top**.** The red and yellow lines indicate the $d_{z^2}$ and $d_{x^2-y^2}/d_{xy}$ orbits. **b,** Photoemission intensity plot at $E_F$ in the $k_x$-$k_z$ plane with $k_y = 0$. **c,** Photoemission intensity along the Γ-A direction. **d,** Band structure projection of Cr $3d_{z^2}$ orbital by first-principal calculations. Inset shows the contrast between calculated band without/with SOC (blue/green line). **e-f,** Schematic of the Cr kagome lattice and the $3d_{z^2}$ orbitals near the Fermi level. **g-i,** Same as **d-f**, but for the $d_{x^2-y^2}$ / $d_{xy}$ orbits. **j**, The abnormal transport anisotropy in $YCr_6Ge_6$. The in-plane/out-of-plane resistivity is indicated by red/blue line. Inset shows the $T^2$ dependence of resistivity at low temperature. **k,** The DOS plot near $E_F$ of Cr $d_{z^2}$ and $d_{x^2-y^2}/d_{xy}$ orbits. **l,** Specific heat measured in zero magnetic field. Inset shows $T^2$ dependence of $C_p/T$ at low temperatures, with the fitting cure $C_p = \gamma_e T + \beta T^3$ ($\gamma_e = 80.5$ mJ/$K^2$ mol, $\beta = 0.3611$ mJ/mol $K^4$).